\documentclass[aps,prl,twocolumn, groupedaddress]{revtex4}
\usepackage{times}
\usepackage{amssymb, amsmath}
\usepackage{graphicx}

\begin{document}

%Version: \today

\title{Time domain measurement of phase noise in a spin torque oscillator}
\author{Mark W. Keller}
\email[]{mark.keller@boulder.nist.gov}
\author{A. B. Kos}
\author{T. J. Silva}
\author{W. H. Rippard}
\author{M. R. Pufall}
\affiliation{National Institute of Standards and Technology, Boulder, CO 80305-3328}

\begin{abstract}
We measure oscillator phase from the zero crossings of the voltage \emph{vs.} time waveform of a spin torque nanocontact oscillating in a vortex mode. The power spectrum of the phase noise varies with Fourier frequency $f$ as $1/f^2$, consistent with frequency fluctuations driven by a thermal source. The linewidth implied by phase noise alone is about 70 \% of that measured using a spectrum analyzer. A phase-locked loop reduces the phase noise for frequencies within its 3 MHz bandwidth.
\end{abstract}

\maketitle

In a spin torque oscillator (STO), dc current transfers spin angular momentum from a thick ``fixed'' ferromagnetic layer to a thin ``free'' ferromagnetic layer. For sufficient current in the appropriate direction, the spin torque counteracts the intrinsic damping torque in the free layer, giving rise to coherent oscillations above a threshold current. As the magnetization precesses along a stable trajectory it produces an oscillating voltage, which in all-metallic devices is due to the giant magnetoresistance effect. The STO frequency can be tuned over a wide range by varying the dc current and a (static) applied magnetic field. Because of their small size ($\sim$ 100 nm), frequency agility, and compatibility with silicon CMOS processing, STOs may be used for applications such as mixing and active phase control in integrated microwave circuits.  Further details about spin transfer torque and STOs can be found in recent reviews \cite{Ralph:2008pr, Rippard:2008pi}.

STOs differ from classical electronic oscillators in several ways, but most important is the essential dependence of frequency on oscillation amplitude \cite{Kim:2008kx}. Because of this, an STO cannot be described by standard circuit models comprising a linear resonator and a feedback amplifier. In particular, Leeson's model for phase noise \cite{Leeson:1966sa} does not apply to STOs. Although oscillator properties such as frequency modulation \cite{Pufall:2005fs} and phase locking \cite{Rippard:2005pd} have been demonstrated in STOs, phase noise has not been directly addressed by previous experiments. The time domain measurement of phase is particularly important because all other quantities that characterize the precision of an oscillator can be derived from it \cite{Stein:1985fi}. When the output voltage of an STO is measured in the frequency domain using a spectrum analyzer, the linewidth is determined by both phase noise and amplitude noise. Although a recent time domain study \cite{Krivorotov:2008oy} provided direct evidence for effects previously inferred from frequency domain data, it did not address phase noise. In this paper we report measurements of phase \emph{vs.} time for an STO, compare the linewidth due to phase noise alone with that measured using a spectrum analyzer, and demonstrate the reduction of STO phase noise using a phase-locked loop.

We studied a nanocontact device from the same batch as those used for a previous study \cite{Pufall:2007hb}. The magnetic layer structure is Ta (3~nm) / Cu (15~nm) / Co$_{90}$Fe$_{10}$ (20~nm) / Cu (4~nm) / Ni$_{80}$Fe$_{20}$ (5~nm) and the region of electrical contact to the layers is a circle of 60~nm nominal diameter. As described previously \cite{Pufall:2007hb, Mistral:2008nr}, spin torque can excite oscillations in these devices that are well below the frequency of uniform ferromagnetic resonance in the extended, continuous film below the contact. Micromagnetic simulations of these modes \cite{Mistral:2008nr} indicate the magnetization has a vortex-like pattern and the oscillations are due to gyrotropic motion of the vortex core about the contact.

\begin{figure}[htbp]
	\includegraphics[scale=0.45]{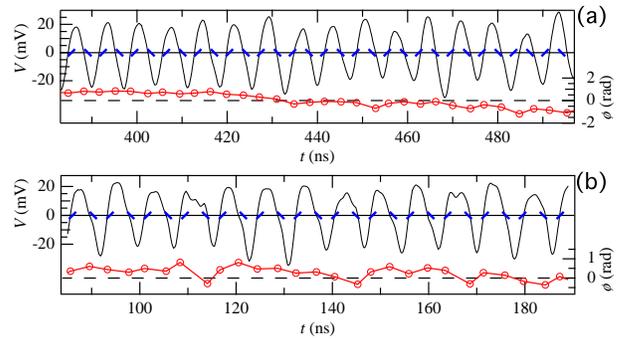}
	\caption{Waveforms and phase from zero crossings for (a) 142.5 MHz mode and (b) 128 MHz mode. The mean value over the entire 4 $\mu$s waveform has been subtracted to remove a dc offset of a few mV. Slash marks show the expected zero crossings (`/' symbol for positive and `$\backslash$' symbol for negative) relative to the first crossing near $t=0$.}
\label{Wave+PhiFig}
\end{figure}

To measure the oscillations, we used a bias tee to separate the dc bias current from the ac response of the device, which we measured using a real-time oscilloscope (1.5 GHz bandwidth and 8 GS/s sampling rate) or a spectrum analyzer. Two stages of preamplification were used to make the ac signal much larger than the input noise of the oscilloscope (the net power gain of 45 dB for the signal path was not subtracted from the data shown here). Harmonics were attenuated by a rolloff in amplifier gain above 150 MHz. As is typical of nanocontact STOs in the vortex mode, our device has several possible modes of oscillation and we adjusted the bias current $I_b$ and applied field $\mu_0 H$ ($\mu_0$ is the magnetic constant) to find modes with linewidth $\leq$ 1 MHz. Data for two modes are presented here. The 142.5 MHz mode used $I_b = $ 6.7 mA and $\mu_0 H =$ 19 mT at 85$^\circ$ from the film plane. The 128 MHz mode used $I_b = $ 6.6 mA and $\mu_0 H =$ 66 mT at 89$^\circ$ from the film plane. (Positive $I_b$ corresponds to electrons flowing from the free layer to the fixed layer.) For both modes, the frequency varied monotonically with $I_b$ at $\approx$ 11 MHz/mA and with $\mu_0 H$ at~$\leq$~1~MHz/mT. All measurements were done at room temperature.

\begin{figure}[bp]
	\includegraphics[scale=0.5]{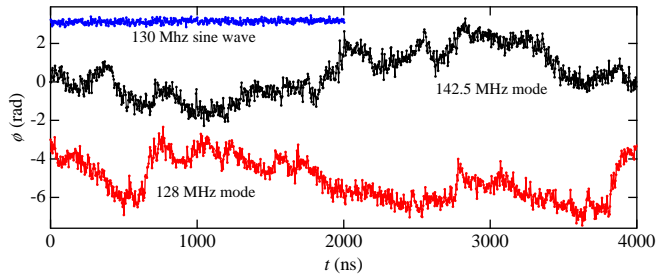}
	\caption{$\phi(t_i)$ from a 4~$\mu$s waveform for each mode (lower curve offset by $-3$~rad for clarity), and from a 2~$\mu$s waveform for a 130~MHz sine wave (offset by 3~rad for clarity)}
\label{PhiAllFig}
\end{figure}

The black traces in Fig. \ref{Wave+PhiFig} show selected subsets of the waveform data. The apparent amplitude variations in these data are primarily due to noise in the amplifiers and do not represent intrinsic amplitude noise in the STO (similar variations were seen for the clean sine wave described below). We calculate the phase of the oscillation at each zero crossing as follows \cite{Cosart:1997if}. The waveform may be written as $V(t) = \left[ {V_0  + \epsilon(t)} \right]\sin \left[ {2\pi \nu_0 t + \phi(t)} \right]$, where $\epsilon(t)$ is the deviation from the nominal amplitude $V_0$, $\nu_0$ is the nominal frequency, and $\phi(t)$ is the deviation from the nominal phase $2\pi \nu_0 t$. Note that $\phi(t)$ includes noise due to changes in precession amplitude combined with the coupling of amplitude and frequency in an STO \cite{Kim:2008kx, Supplement1}. Zero crossings of $V(t)$ occur when $2\pi \nu _0 t + \phi (t) = n \pi$, where $n$ is an even (odd) integer for crossings with a positive (negative) slope. The set of values $\{n_i, t_i\}$ for a waveform gives measurements of phase deviation at discrete times: $\phi (t_i ) = 2\pi ( n_i / 2 - \nu_0 t_i )$. For $\nu_0$ we use the mean frequency determined from the total number of zero crossings during the entire waveform. Returning to Fig. \ref{Wave+PhiFig}, the slash marks along the line $V = 0$ show the expected zero crossings (relative to the first crossing near $t = 0$) \cite{ZeroCrossAsymmNote2} and the lower curve shows $\phi(t_i)$. Figure \ref{PhiAllFig} shows $\phi(t_i)$ for the entire span of 4~$\mu$s acquired with the oscilloscope. The noise floor of our measurement is illustrated by $\phi(t_i)$ for a clean 130~MHz sine wave from a signal generator with negligible phase noise. This signal was attenuated to have the same amplitude as the raw STO signal and measured after passing through the same signal path.

The time domain data in Figs.~\ref{Wave+PhiFig} and \ref{PhiAllFig} illustrate some typical characteristics of the phase noise for the vortex modes considered here. Fig.~\ref{Wave+PhiFig}a shows an example of several periods with $\phi$ nearly constant, then a relatively gradual change of about $-1.5$~rad between $t$ = 420~ns and 490~ns. Fig.~\ref{Wave+PhiFig}b shows more abrupt changes in $\phi$, including multiple changes within a half period that cannot be detected by our zero crossing method (near $t$ = 110~ns and 143~ns).  Figure~\ref{PhiAllFig} shows the random walk character of the phase variations over long times, with an rms fluctuation over 4~$\mu$s of 1.2~rad for the 142.5~MHz mode and 1.1~rad for the 128~MHz mode. The rms fluctuation for the sine wave over 2~$\mu$s is 0.10~rad.

\begin{figure}[tbp]
	\includegraphics[scale=0.5]{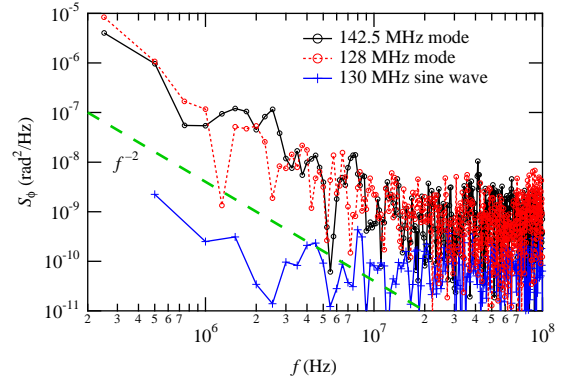}
	\caption{Power spectral density of phase fluctuations for the data in Fig. \ref{PhiAllFig}. Dashed line is a guide to the eye showing $1/f^2$ dependence.}
\label{SphiFig}
\end{figure}

Figure \ref{SphiFig} shows the one-sided power spectral density (PSD) of $\phi(t_i)$ \cite{Supplement1}, which falls off as $1/f^2$ until it saturates at $(1.0 \pm 0.07) \times 10^{-9}$~rad$^2$/Hz for both modes. (All uncertainties in this paper are one standard deviation.) The PSD for the sine wave saturates at $(9.4 \pm 0.6) \times 10^{-11}$~rad$^2$/Hz, which is consistent with the noise floor for zero crossings set by the Johnson noise of a 50~$\Omega$ impedance propagated through the signal path \cite{Supplement1}. The saturation of the STO phase noise at a higher value could be explained by additional white noise on $V(t)$ (before amplification) of about 2~nV$/$Hz$^{1/2}$, a level too small to measure directly in our setup. The phase added along the signal path varies slightly with frequency, but the noise floor due to this effect is $\leq 10^{-11}$~rad$^2$/Hz.

In the equation for $V(t)$, the total phase $\theta (t) \equiv 2\pi \nu_0 t + \phi(t)$ defines an instantaneous frequency $\nu (t) \equiv  (1/2\pi)d\theta / dt = \nu_0 + (1/2\pi)d\phi / dt$. Phase noise and frequency noise are equivalent ways of expressing the instability of an oscillator \cite{Stein:1985fi}, and because frequency is the derivative of phase, their PSDs are related by $S_\nu  (f) = f^2 S_\phi  (f)$. The data in Fig.~\ref{SphiFig} thus imply that $S_\nu$ is independent of $f$ (below $10^7$~Hz), suggesting that frequency fluctuations in our vortex mode STO are driven by a white noise source such as thermal fluctuations. The mean value of $S_{\nu}$ for $f \leq 10^7$~Hz is $(2.3 \pm 0.3) \times 10^5$~Hz$^2$/Hz for the 142.5~MHz mode and $(1.7 \pm 0.2) \times 10^5$~Hz$^2$/Hz for the 128~MHz mode.

\begin{figure}[tbp]
	\includegraphics[scale=0.4]{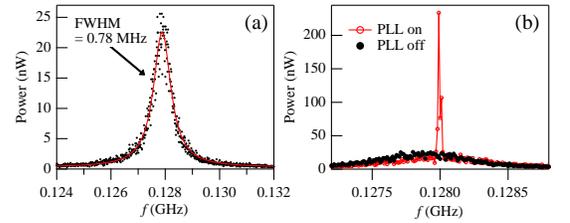}
	\caption{Spectrum analyzer measurements of amplified STO signal for the 128~MHz mode (resolution bandwidth~=~10~kHz). (a) PLL off, with Lorentzian fit. (b) PLL on, with same data as in (a) for comparison.}
\label{SvFig}
\end{figure}

The value of $S_\nu$ allows us to calculate the contribution of phase noise to linewidth as follows \cite{Blaquiere:1966ng}. Neglecting amplitude noise, the ensemble-averaged $V(t)$ for oscillators whose phases undergo a diffusive random walk is a damped sinusoid. The power spectrum of $V(t)$ (the quantity measured by a spectrum analyzer) has a full-width-at-half-maximum $\Delta \nu$ that is proportional to the PSD of the white frequency noise: $\Delta \nu = \pi S_{\nu}$ \cite{Blaquiere:1966ng}, where $S_{\nu}$ is the one-sided PSD \cite{Supplement1}. Thus our values of $S_{\nu}$ above imply linewidths of $(0.72 \pm 0.10)$~MHz for the 142.5~MHz mode and $(0.54 \pm 0.08)$~MHz for the 128~MHz mode. Using a spectrum analyzer, we measured linewidths of $(1.05 \pm 0.14)$~MHz for the 142.5~MHz mode and $(0.78 \pm 0.11)$~MHz for the 128~MHz mode. We conclude that phase noise is responsible for about 70 \% of the linewidth in both cases, with the rest presumably due to amplitude noise. (As mentioned above, STO amplitude noise is masked by amplifier noise in the waveform data presented here.)

A common way to improve the phase noise of an oscillator is to use a phase-locked loop (PLL). Although abrupt phase jumps such as those seen in Fig. \ref{Wave+PhiFig}b cannot be removed, a PLL can stabilize the STO against slower phase variations such as those in Fig. \ref{Wave+PhiFig}a. We built a PLL in which $V(t)$ from both the STO and a signal generator at the reference frequency are converted into square waves using digital comparators. Timing jitter between these square waves produces an error signal that is fed into an integrator, and the integrated error signal is added to the dc bias current of the STO using a dual-input bias tee. Distortion in the digital comparators becomes excessive above 200 MHz, and the overall feedback bandwidth is limited to 3 MHz by the input of the bias tee.

\begin{figure}[tbp]
	\includegraphics[scale=0.5]{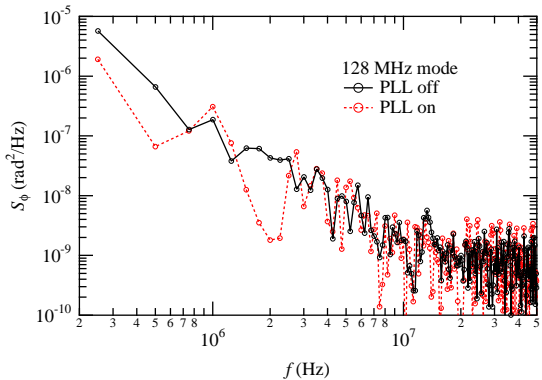}
	\caption{Power spectral density of phase fluctuations with PLL off and on. Dashed line is a guide to the eye showing $1/f^2$ dependence.}
\label{SphiPLLFig}
\end{figure}

The effect of the PLL as measured by the spectrum analyzer is shown in Fig. \ref{SvFig}. The left plot is the spectrum of the 128 MHz mode before locking the PLL, as well as a Lorentzian fit giving $\Delta \nu = 0.78$~MHz. The right plot is the spectrum with the PLL on, showing a large increase in power at the 128~MHz reference frequency. Comparing the area of the narrow peak with the total area over a 6~MHz span, we find that 20 \% of the power occurs within $\pm 1$ resolution bandwidth of the reference frequency, for resolution bandwidths as low as 10 Hz. We obtained a similar result for the 142.5~MHz mode.

The effect of the PLL on $S_\phi$ is shown in Fig.~\ref{SphiPLLFig}, where one can see a modest decrease in noise for frequencies below the 3~MHz limit mentioned above. The variance obtained by integrating $S_\phi$ over $f$ is 1.2~rad$^2$ with the PLL off and 0.53~rad$^2$ with the PLL on. The effect of the PLL is more dramatic in Fig.~\ref{SvFig} than in Fig.~\ref{SphiPLLFig} because the spectrum analyzer spends $\sim 100~\mu$s at each frequency while acquiring the spectrum, whereas the waveforms used to obtain $S_\phi$ last only 4~$\mu$s. Longer waveforms (using an oscilloscope with more memory) should reveal a more significant decrease in $S_\phi$ at lower frequencies.

The technique described here for directly measuring phase and its power spectral density offers a valuable characterization tool for STOs. It gives information that is not available from a linewidth measurement and it can discriminate among models for noise that predict similar linewidths but different forms for $S_\phi (f)$. Such tests will be useful in understanding the behavior of STOs and in pushing their performance toward that required by applications.

\newpage
\newpage
\appendix{\large \textbf{Supplementary Material (EPAPS)} }
\section*{Coupling between amplitude and phase}

{\large In a linear oscillator, perturbations that change the phase
but do not cause the oscillator to deviate from its stable trajectory,
}\emph{\large i.e.}{\large , perturbations }\emph{\large along}{\large{}
the stable trajectory, cause phase noise only. Similarly, perturbations
}\emph{\large transverse}{\large{} to the stable trajectory cause amplitude
noise only, since the frequency of oscillation is independent of amplitude.
Unlike in a linear oscillator, the frequency of an STO depends on
the amplitude of oscillation \cite{a_Rezende:2005vl,a_Slavin:2005cs},
with important consequences for phase noise \cite{a_Kim:2008kx}. During
the time the STO is perturbed away from its stable trajectory, the
magnetization moves at a different rate, so when it returns to the
stable trajectory it will be either advanced or retarded relative
to its unperturbed motion. As a consequence, deviations from the stable
trajectory result in both amplitude and phase noise. In the equation
for $V(t)$, both $\epsilon(t)$ and $\phi(t)$ are affected by transverse
perturbations, but it is sufficient to measure either one to detect
these perturbations.}{\large \par}

\section*{Power spectral density}

{\large Computation of the Fourier transform of $\phi(t)$ is slightly
complicated by the fact the the times $t_{i}$ at which $\phi$ is
measured are not uniformly spaced. For oscillators with low phase
noise it is appropriate to proceed by assuming that the phase change
within one period is small (<\textcompwordmark{}< 1 rad) and then
interpolating from the actual times $t_{i}$ to uniformly spaced times.
This assumption is clearly not valid for our waveforms (see Fig. 1
of the paper). We therefore used the Lomb periodogram described in
\cite{a_Press:2007ek}, which is designed to handle nonuniform data.
As is common for real signals \cite{a_Press:2007ek}, we define the
PSD over positive frequencies only and normalize so that its integral
over these frequencies gives the variance of the waveform. With this
definition, the PSD at each value of $f$ is twice as large as for
the two-sided spectrum often used in theoretical work, and we took
this into account when applying the results of Blaquiere for the relation
between $f$ and $S_{\nu}$. We multiplied $\phi(t)$ by a Hann window
before computing the PSD in order to remove spurious effects due to
the non-periodic nature of the data, and the effect of this window
was included in the normalization of the PSD \cite{a_Press:2007ek}.}{\large \par}

\section*{Noise floor}

{\large The output of an oscillator having no phase noise but with
an additive noise voltage $\delta v$ is\begin{equation}
V(t)=V_{0}\sin(2\pi\nu_{0}t)+\delta v.\end{equation}
At it zero crossings, $V(t)$ has a slope of\begin{equation}
\frac{dV}{dt}=\pm2\pi\nu_{0}V_{0},\end{equation}
A weak voltage noise $\delta v<<V_{0}$ will cause a timing noise
given approximately by\begin{equation}
\delta t\approx\frac{\delta v}{2\pi\nu_{0}V_{0}},\end{equation}
 which implies a phase noise\begin{equation}
\delta\phi=2\pi\nu_{0}\delta t\approx\frac{\delta v}{V_{0}}.\end{equation}
The same result is derived in a slightly different way in \cite{a_Baghdady:1965vc}
(see equations 30 through 34). This result implies that the PSD of
phase noise due to additive voltage noise is\begin{equation}
S_{\phi}=\frac{S_{v}}{V_{0}^{2}}.\label{eq:SphiFloor}\end{equation}
We note that Eq. \ref{eq:SphiFloor} differs by a factor of two from
the often-cited result by Leeson in \cite{a_Leeson:1966sa}.}{\large \par}

{\large The noise in our signal path is nearly independent of the
impedance seen by the input of the first amplifier for loads less
than 50 $\Omega$. Therefore, we estimate the minimum $S_{v}$ presented
to the first amplifier as that of a 50 $\Omega$ load at 290 K. Using
the gains and noise temperatures of our amplifiers, we propagate this
noise through our 50 $\Omega$ signal path in the standard way \cite{a_Pozar:2005ei}
and find an ouput noise of $S_{v}=8.1\times10^{-14}\mathrm{\; V^{2}/Hz}$.
Combining this with $V_{0}=$ 22 mV gives $S_{\phi}=1.7\times10^{-10}\;\mathrm{rad}^{2}\mathrm{/Hz}$
(with an uncertainty of about a factor of 2) as the noise floor due
to Johnson noise in our measurement.}{\large \par}


\begin{thebibliography}{10}
\bibitem{ZeroCrossAsymmNote2}
 The strong harmonic content of vortex mode oscillations makes our waveforms
  asymmetric, with the peaks 10 \% to 20 \% wider than the valleys. The
  expected zero crossings include this asymmetry to avoid a systematic offset
  in phase between positive and negative crossings.
\end{thebibliography}
\end{document}